\documentclass[a4paper,12pt]{article}
\usepackage{amsfonts,amsthm,amsmath,amssymb,graphicx,hyperref,youngtab}

\def\g{\gamma}
\def\G{\Gamma}

\def\s{\sigma}

\newcommand{\cP}{\mathcal P}

\newcommand{\be}{\begin{equation}}
\newcommand{\bea}{\begin{eqnarray}}
\newcommand{\ee}{\end{equation}}
\newcommand{\eea}{\end{eqnarray}}

\def\g{\gamma}
\def\s{ \sigma}

\begin{document}

\makeatletter
\@addtoreset{equation}{section}
\makeatother
\renewcommand{\theequation}{\thesection.\arabic{equation}}

\rightline{WITS-CTP-110}
\vspace{1.8truecm}

\vspace{15pt}


{\LARGE{  
\centerline{   \bf A basis for large operators in N=4 SYM} 
\centerline  {\bf  with orthogonal gauge group} 
}}  

\vskip.5cm 

\thispagestyle{empty} \centerline{
    {\large \bf Pawel Caputa$^a$\footnote{{\tt pawel.caputa@wits.ac.za}}, Robert de Mello Koch$^{a,b}$\footnote{{\tt robert@neo.phys.wits.ac.za}} }
   {\large \bf and Pablo Diaz$^a$\footnote{{\tt Pablo.DiazBenito@wits.ac.za}} }}

\vspace{.4cm}
\centerline{{\it $^{a}$National Institute for Theoretical Physics,}}
\centerline{{\it Department of Physics and Centre for Theoretical Physics,}}
\centerline{{\it University of Witwatersrand, Wits, 2050, } }
\centerline{{\it South Africa } }
\vspace{.4cm}
\centerline{{\it $^{b}$Institute of Advanced Study,}}
\centerline{{\it Durham University}}
\centerline{{\it Durham DH1 3RL, UK}}
\vspace{.4cm}

\vspace{1.4truecm}

\thispagestyle{empty}

\centerline{\bf ABSTRACT}

\vskip.4cm 

We develop techniques to study the correlation functions of ``large operators'' whose bare dimension grows parametrically with $N$,
in $SO(N)$ gauge theory. 
We build the operators from a single complex matrix.
For these operators, the large $N$ limit of correlation functions is not captured by summing only the planar diagrams.
By employing group representation theory we are able to define local operators which generalize the Schur polynomials of the 
theory with gauge group $U(N)$. 
We compute the two point function of our operators exactly in the free field limit showing that they diagonalize the two point function.
We explain how these results can be used to obtain the exact free field answers for correlators of operators in the trace basis.

\setcounter{page}{0}
\setcounter{tocdepth}{2}

\newpage

\tableofcontents

\setcounter{footnote}{0}

\linespread{1.1}
\parskip 4pt

{}~
{}~

\section{Introduction}

In this article we apply group representation theory methods to organise local operators built using a single complex matrix
and to compute their two point correlation functions in $SO(N)$ gauge theory when $N$ is finite. 
Using the AdS/CFT duality\cite{malda,Gubser:1998bc,Witten:1998qj}, probing finite $N$ physics of the gauge 
theory\cite{Balasubramanian:2001nh} corresponds to the study of non-perturbative objects such as giant graviton 
branes\cite{mst,myers,hash}, as well as aspects of spacetime geometry captured in the stringy exclusion 
principle\cite{Maldacena:1998bw}, that take us beyond the supergravity approximation.
This programme was initiated by Corley, Jevicki and Ramgoolam in \cite{cjr} for the half-BPS operators constructed using a single 
complex matrix. Progress on the study of the finite $N$ physics has been rapid and we now know of a number of bases of local
operators that diagonalize the free field two point function\cite{dssi,Kimura:2007wy,BHR1,BHR2,Bhattacharyya:2008rb,Kimura:2009jf,Kimura:2012hp}
and we know how to diagonalize the one-loop dilatation operator\cite{Koch:2010gp,DeComarmond:2010ie} for certain operators dual to giant graviton 
branes\cite{Carlson:2011hy,gs,Koch:2011hb,DCI}. 
This diagonalization has provided new integrable sectors, with the spectrum of the dilatation operator reducing to that of 
decoupled harmonic oscillators which describe the excitations of the strings\cite{gs,Koch:2011hb,DCI} attached to the giants. 
Integrability in the planar limit was discovered in \cite{mz,bks} and is reviewed in \cite{intreview}.

All work to date has focused on field theories with gauge group $U(N)$ or $SU(N)$\footnote{See \cite{gwyn} for a study
of the $SU(N)$ theory.}. 
There are good reasons to extend these studies to field theories with $SO(N)$ or $Sp(N)$ gauge group. 
In this work we will focus on the extension to $SO(N)$ gauge groups leaving the $Sp(N)$ gauge group for the future.
A detailed study of the spectral problem of ${\cal N}=4$ super Yang-Mills with these gauge 
groups has been carried out in \cite{Caputa:2010ep}.
At the planar level, the essential difference between the theories with gauge groups $U(N)$
or $SO(N)$ is that in the $SO(N)$ case certain states are projected out.
Thus, the planar spectral problem of the $SO(N)$ theory can again be mapped to an
integrable spin chain\cite{Caputa:2010ep}. 
At the non-planar level there are genuine differences and the leading non-planar corrections come from
ribbon graphs that triangulate non-orientable Feynman diagrams with a single cross-cap.
These corrections produce a non-local spin chain interaction which removes a section of the
chain, reverses its orientation and then reinserts it\cite{Caputa:2010ep}.
As mentioned above, we are interested in developing techniques that allow us to study ``large operators'' whose bare dimension
grows parametrically with $N$. 
In this case, the large $N$ limit of correlation function is not captured by summing only the planar 
diagrams\cite{Balasubramanian:2001nh}, so that correlation functions of large 
operators are sensitive to the non-planar structure of the theory.
Given the clear differences at the non-planar level between the theories with gauge groups $U(N)$
or $SO(N)$, we are sure to learn something new by studying large operators in the theory with $SO(N)$ gauge group. 
Apart from this field theoretic motivation, it is known that the ${\cal N}=4$ super Yang-Mills with $SO(N)$ gauge 
group is AdS/CFT dual to the AdS$_5\times {\mathcal R}$P$_5$ geometry\cite{Witten:1998xy}.
In this case one expects a non-oriented string theory so that the study of non-perturbative stringy physics,
which is captured by the finite $N$ physics of the gauge theory, is likely to provide new insights extending
what can be learned from the AdS$_5\times$S$^5$ example which involves oriented string.
For studies in this direction see \cite{Aharony:2002nd}. 

It is useful to review how the inner product is diagonalized for gauge group $U(N)$, stressing those features that will
need to be considered to achieve the desired generalization to gauge group $SO(N)$. 
For the $U(N)$ gauge theory, thanks to the global $U(N)$ symmetry, we know that the two point function of the 
hermitian adjoint Higgs fields $X_{ij}$ and $X_{kl}$ is only non-zero if $i=l$ and $j=k$. 
Given this fact, it makes sense to recognize that the row and column indices of $X$ belong to different vector 
spaces and reflect this in our notation by writing $X^i_j$ and $X^k_l$.
We then use the symmetric group to organize the row and column indices separately. 
To obtain gauge invariant observables we need to take a trace.
Since the free field theory Wick contractions can be realized as a sum over permutations, this organization 
diagonalizes the Zamalodchikov inner product of the conformal field theory provided by the two point function.
For the gauge group $SO(N)$ the structure that emerges is rather different.
The two point function of $X_{ij}$ and $X_{kl}$ is non zero when $i=l$ and $j=k$ or if $i=k$ and $j=l$.
Thus, we must recognize that the row and column indices belong to the same vector space.
This implies that our organization of operators in the theory has a different structure as compared to the $U(N)$ case.
In section three we study the simplest version of this problem, provided by the $O(N)$ vector model.
This toy model problem is ideal as it displays all the elements of the general problem.
We start by considering a product of vectors which, since they are left uncontracted, define an $SO(N)$ tensor.
The indices of this tensor can be organized using permutations, thereby breaking it into pieces that do not
mix under the action of the symmetric group.
The free field Wick contractions are represented in this language, by a projection operator projecting onto the one 
dimensional irreducible representation (irrep) of the symmetric group labeled by a Young diagram with a single row.
By contracting indices in pairs we are able to extract a gauge invariant operator from the $SO(N)$ tensor.
We compute the two point functions in this model exactly, in the free field limit(see Appendix A
for all the details).
The logic of this vector model computation plays a central role in the computation of two 
point functions in the $SO(N)$ gauge theory.

In section four, using the lessons gained from the toy model, we study the $SO(N)$ gauge theory. 
We start by using the adjoint Higgs fields to define an $SO(N)$ tensor.
The indices of this tensor can again be organized using permutations, which breaks it into pieces that do not
mix under the action of the symmetric group. 
Each such piece is labeled by a Young diagram $R$.
A non-trivial difference compared to the toy model, is that now the complete set of Wick contractions
are given by a sum over the wreath product $S_n[S_2]$.
This sum again defines a projector, but now we project onto the one dimensional antisymmetric irrep of $S_n[S_2]$.
We then introduce a complete basis of local operators that diagonalizes the two point function, by
identifying which representations $R$ subduce a copy of the antisymmetric irrep of $S_n[S_2]$.
It turns out that $R$ must have both an even number of columns and an even number of rows.
We are able to give a simple closed formula for the two point function of our operators.
In obtaining these results, we need to compute a matrix element defined using particular representations 
of the wreath group.
Remarkably, this matrix element has been computed in \cite{ivanov}.
We review the relevant results of \cite{ivanov} in an Appendix.

In section 5 we show how our results can be used to obtain correlation functions of operators in the trace basis, 
to all orders in $1/N$. 
This again generalizes results that are known for the $U(N)$ gauge theory\cite{Corley:2002mj,Caputa:2012dg}.

$SO(N)$ has two basic invariant tensors, $\delta^{ij}$ and $\epsilon^{i_1 i_2 \cdots i_N}$.
We can extract gauge invariant operators from an $SO(N)$ tensor by contracting indices in pairs using $\delta^{ij}$s 
or by contracting $N$ indices at a time using $\epsilon^{i_1 i_2 \cdots i_N}$.
We pursue the construction of gauge invariant operators constructed using $\epsilon^{i_1 i_2 \cdots i_N}$ in section 6.
We find new operators, including the Pfaffian and compute its two point function using our technology. 

Section 7 is used for a discussion of our results.

\section{Notation and Comments}

We will often use the space $V^{\otimes q}$, which is defined to be the space of $SO(N)$ tensors with $q$ indices. 
Our notation for this space simply reflects that it is isomorphic to the space obtained by taking the tensor 
product of $q$ copies of the $N$ dimensional space $V$ that carries the vector representation of $SO(N)$.
$V^{\otimes q}$ is a vector space of dimension $N^q$. 
We will denote the states of this space as the kets $|i_1 i_2 \cdots i_q\rangle$. 
There is a natural action of $\sigma\in S_q$ on this space $\sigma: V^{\otimes q}\to V^{\otimes q}$ which plays
a central role in what follows. 
Thinking of $\sigma$ as a matrix, it has matrix elements
\bea
   \langle j_1 j_2\cdots j_q|\sigma |i_1 i_2 \cdots i_q\rangle
    =\delta^{j_1}_{i_{\sigma (1)}}\delta^{j_2}_{i_{\sigma (2)}}\cdots\delta^{j_q}_{i_{\sigma (q)}}
   \label{actonspace}
\eea

If we consider a permutation $\sigma$ acting on $V^{\otimes q}$, with the action given in (\ref{actonspace}), we simply write it as $\sigma$.
We will also sometimes need to consider $\sigma$ as an operator in the carrier space of some irreducible representation of the symmetric group 
specified by Young diagram $R$, with $R$ a partition of $q$.
In this case we write the matrix representing permutation $\sigma$ as $\Gamma^R (\sigma)$. 
$\chi_R (\sigma)$ denotes the character of $\sigma$ in representation $R$.

We use the standard notation $R\vdash q$ to specify that $R$ is a partition of $q$.

The wreath product $S_n[S_2]$ will play an important role in what follows. 
It is the semi-direct product $(S_2)^n\rtimes S_n$.
A particularly useful reference for the representation theory for the wreath product is \cite{jameskerber}.

We will make heavy use of projection operators in what follows. 
The action of $\sigma$ on $V^{\otimes q}$ is reducible. 
The operator ($R\vdash q$)
\bea
  P_{V^{\otimes q}\to R}={1\over q!}\sum_{\sigma\in S_q}\chi_R (\sigma) \sigma
\eea
projects from $V^{\otimes q}$ to the subspace that carries irrep $R$ of $S_q$. In general, there is more
than one copy of $R$ in $V^{\otimes q}$; we will not need to track or resolve this multiplicity.
$P_{V^{\otimes q}\to R}$ is not normalized correctly to qualify for the name projection operator, since
\bea
P_{V^{\otimes q}\to R} P_{V^{\otimes q}\to S} = {\delta_{RS}\over d_R}P_{V^{\otimes q}\to R}
\eea
With a slight abuse of language, we will refer to $P_{V^{\otimes q}\to R}$ as a projection operator.
The operators
\bea
  P_{R\to [A]}={1\over 2^q q!}\sum_{\sigma\in S_q[S_2]}\chi_{[A]} (\sigma) \Gamma^R(\sigma)
\eea
with $[A]$ some irreducible representation of $S_q[S_2]$, projects from the carrier space of $R$ an irrep of $S_{2q}$ to
the carrier space of $[A]$. 
Our convention is to refer to representations of the symmetric group with a capital letter 
and to refer to representations of the wreath product $S_q[S_2]$ with a capital letter in square braces.
Again, with a slight abuse of notation we will refer to $P_{R\to [A]}$ as a projection operator.

To further simplify the notation we will denote all projection operators projecting from $V^{\otimes q}$ with
a curly $\cP$ and projectors projecting from a carrier space of a symmetric group representation with a normal $P$
omitting the domain of the operator in both cases.
Thus, $P_{V^{\otimes q}\to R}$ is denoted $\cP_R$ while $P_{R\to [A]}$ is denoted $P_{[A]}$.

\section{Toy Model Problem}

To start we'll study a simple problem, that will allow us to develop the ideas that are useful for the $SO(N)$ gauge theory.
The simple problem is a free $O(N)$ vector model, with 2 flavors in 0 dimensions.
Denote the two flavors $x^i$ and $y^i$ with color index $i=1,2,...,N$.
The non-zero Wick contractions are
\bea
  \langle x^i x^j\rangle = \delta^{ij} = \langle y^i y^j \rangle
\eea
There is an $O(N)$ symmetry which acts as
\bea
  M:(x^i,y^j)\to ((Mx)^i,(My)^j) \qquad M\in {\rm SO}(N)
\eea
Ultimately we will think of this $O(N)$ symmetry as a gauge symmetry so that physical observables are $O(N)$ singlets.

Introduce the notation $z^i = x^i + i y^i$ and $z_i = x^i - i y^i$. The Wick contractions are
\bea
  \langle z_i z_j\rangle = 0 = \langle z^i z^j \rangle\qquad \langle z^i z_j\rangle = \delta^i_j
\eea

Study the operators
\bea
  T^{i_1 i_2 \cdots i_n}=z^{i_1} z^{i_2}\cdots z^{i_n} \in V^{\otimes n}
\eea
and
\bea
  (T^{i_1 i_2 \cdots i_n})^*=T_{i_1 i_2 \cdots i_n}=z_{i_1} z_{i_2}\cdots z_{i_n} 
\eea
We will often employ a short hand, collecting the $n$ indices $i_1 i_2 \cdots i_n$ into a single index $I$.
The action of the symmetric group on $V^{\otimes\, n}$ gives us a particularly useful language with which we can discuss Wick contractions.
Indeed
\bea
  \langle T^{i_1 i_2 \cdots i_n}T_{j_1 j_2 \cdots j_n}\rangle =\sum_{\sigma\in S_n}\sigma^I_J
\eea
There is also an action of $U(N)$ on $V^{\otimes n}$ which commutes with $S_n$\footnote{This $U(N)$ is not a symmetry of the theory.
An $SO(N)$ subgroup is a symmetry. Nevertheless, $U(N)$ makes an appearance because it is the centralizer of the symmetric group and it 
is for this reason that our operators are labeled with Young diagrams which are more naturally related to $U(N)$ than to $SO(N)$}. 
As a consequence of these commuting actions
we have Schur-Weyl duality which links the group theory of the symmetric and unitary groups. Using this duality we can 
decompose the tensor $T^{i_1 i_2 \cdots i_n}$ into components which have orthogonal two point function. The projectors
\bea
  (\cP_R)^I_J ={1\over n!}\sum_{\sigma\in S_n}\chi_R(\sigma)\sigma^I_J
  \label{projectors}
\eea
commute with $\sigma$ and are orthogonal
\bea
  \left[ \cP_R,\sigma\right]=0\quad \forall\sigma\in S_n \cr
   \cP_R \cP_S \propto \delta_{RS}
\eea
The labels $R$ are Young diagrams built using $n$ boxes. Consider the new operators
\bea
  (T_R)^K = (\cP_R)^K_I T^I \qquad (T_S)_L=(\cP_S)^J_L T_J
\eea
It is clear that
\bea
  \langle (T_R)^K (T_S)_L \rangle =\sum_{\sigma\in S_n}(\cP_R \sigma \cP_S)^K_L
=\sum_{\sigma\in S_n}(\sigma \cP_R \cP_S)^K_L\propto \delta_{RS}
\eea
This is the first key idea: by representing the sum of Wick contractions as a permutation operator on $V^{\otimes n}$ we can diagonalize
the two point function by constructing operators that commute with permutations. This problem has already been solved: the operators we
need are just the projectors given in (\ref{projectors}).

To proceed further, we need to evaluate the sum
\bea
  \sum_{\sigma\in S_n}(\sigma \cP_R \cP_S)^K_L
\eea
The key observation which allows the evaluation of this sum is that
\bea
  \sum_{\sigma\in S_n}\sigma^I_J = n! \cP_{(n)}
\eea
where $(n)$ is the Young diagram that has a single row of $n$ boxes. This follows because $\chi_{(n)}(\sigma )=1$ for all $\sigma$.
It is now clear that 
\bea
  \langle (T_R)^K (T_S)_L \rangle  \propto \delta_{RS}\delta_{R (n)}
\eea
Thus, we have a single operator $(T_{(n)})^K$. To get a physical observable, we need to contract all the indices in pairs.
This forces us to consider even $n$, i.e. $n=2p$ so that the physical observable is
\bea
  O_{(n)}=(T_{(n)})^{i_1 i_1 i_2 i_2 \cdots i_p i_p} = (z^i z^i)^p
\eea
We knew that this is the only $O(N)$ invariant we can build from a single vector before we even started. 
However, we now have a strategy for the construction of operators that diagonalize the two point 
function that will generalize to the $SO(N)$ gauge theory.
Concretely, here is a summary of the logic
\begin{itemize}
\item[1.] Recognize the sum over Wick contractions as a sum over permutations in $V^{\otimes \, n}$.
\item[2.] Build operators that diagonalize the free field two point functions by utilizing orthogonal projectors
          that commute with all permutations and hence, in particular, the Wick contractions.
          This approach has the advantage that the projectors are well known from symmetric group theory.
\item[3.] From these diagonal operators extract the gauge invariant operators.
\end{itemize}
The exact two point function can now be computed by evaluating
\bea
\langle O_{(2p)}O_{(2p)}^*\rangle =  \sum_{\sigma\in S_{2p}}\chi_{(2p)}(\sigma)\sigma^{i_1 i_1 \cdots i_p i_p}_{j_1 j_1 \cdots j_p j_p}
\eea
We perform this evaluation in detail in the Appendix \ref{detailed} and verify 
the computation by solving Schwinger-Dyson equations.
The result is
\bea
  \langle O_{(2p)}O_{(2p)}^*\rangle = \langle (\vec{z}\cdot \vec{z})^p(\vec{\bar{z}}\cdot \vec{\bar{z}})^p\rangle 
                                  = 2^p p!\prod_{i=0}^{p-1} (N+2i)
\label{exactvector}
\eea
This can also be written as
\begin{equation}
 \langle O_{(2p)}O_{(2p)}^*\rangle = 2^{2p}p! \frac{\Gamma(N/2+p)}{\Gamma(N/2)}
\end{equation}

\section{ $SO(N)$ gauge theory } 

Consider an $SO(N)$ gauge theory with two species of matrices $X^{ab}$ and $Y^{ab}$, $a,b=1,...,N$.
The non-zero two point functions are
\bea
  \langle X^{ab}X^{cd}\rangle = \delta^{ad}\delta^{bc} -\delta^{ac}\delta^{bd}= \langle Y^{ab}Y^{cd}\rangle
\eea
Introduce the complex combinations
\bea
  Z^{ab}=X^{ab}+iY^{ab}\qquad Z_{ba}=X^{ab}-iY^{ab}
\eea
The non-zero two point function is now
\bea
  \langle Z^{ab} Z_{dc}\rangle = \delta^a_d\delta^b_c -\delta^a_c\delta^b_d
\eea

Introduce the tensor
\bea
  T^{a_{1,1} a_{1,2} a_{2,1} a_{2,2} a_{3,1} \cdots a_{n,1}a_{n,2}}=Z^{a_{1,1} a_{1,2}} Z^{a_{2,1} a_{2,2}}\cdots Z^{a_{n,1} a_{n,2}}
\eea
which lives in $V^{\otimes 2n}$. 
This labelling of the $2n$ indices of $T$ will prove to be very useful when we come to describe the Wick contractions.
We will also use
\bea
  T_{a_{1,1} a_{1,2} a_{2,1} a_{2,2} a_{3,1} \cdots a_{n,1}a_{n,2}}=Z_{a_{1,1} a_{1,2}} Z_{a_{2,1} a_{2,2}}\cdots Z_{a_{n,1} a_{n,2}}
\eea
To repeat the lesson we learned in the last section,
we need to realize the Wick contractions as a sum over permutations in $V^{\otimes 2n}$. 
This time the set of Wick contractions is not
a sum over the symmetric group, but rather it is a sum over the wreath product $S_n [S_2]$. 
The $S_n$ factor will act on the first index ($i$) of the subscript $a_{ij}$ while the $S_2$ 
factor will act on the second index ($j$).
Since the Wick contractions are again given as a sum over permutations, it makes sense to again
decompose $T^{a_{1,1} a_{1,2} a_{2,1} a_{2,2} a_{3,1} \cdots a_{n,1}a_{n,2}}$ into irreducible 
components that don't mix under the action of the symmetric group $S_{2n}$. Since the matrix
$Z$ itself is antisymmetric we know that
\bea
   T^{a_1 a_2} {\rm \,\,\, is\,\,\, in\,\,\, the\,\,\, } \yng(1,1) {\rm \,\,\, representation}
\eea
\bea
   T^{a_{1,1} a_{1,2} a_{2,1} a_{2,2}} {\rm \,\,\, is\,\,\, in\,\,\, the\,\,\, } 
      \yng(1,1)\otimes \yng(1,1) = \yng(2,2)\oplus\yng(2,1,1)\oplus\yng(1,1,1,1) {\rm \,\,\, representation}
\eea
\bea
  T^{a_{1,1} a_{1,2} a_{2,1} a_{2,2} a_{3,1} a_{3,2}} {\rm \,\,\, is\,\,\, in\,\,\, the\,\,\, } 
      \yng(1,1)\otimes \yng(1,1)\otimes \yng(1,1)\\
 = 3\yng(2,2,1,1)\oplus 2\yng(2,1,1,1,1)\oplus\yng(1,1,1,1,1,1)\oplus
      \yng(3,3)\oplus 2\yng(3,2,1)\oplus\yng(3,1,1,1)\oplus\yng(2,2,2) {\rm \,\,\, representation}
\eea
etc. So it is again easy to decompose the tensor 
$T^{a_{1,1} a_{1,2} a_{2,1} a_{2,2} a_{3,1} \cdots a_{n,1}a_{n,2}}$
into components that don't mix when we compute their two point function.
The basic formula is
\bea
  (T_R)^A = (\cP_R)^A_B T^B
  \label{basicformula}
\eea
where the projector $\cP_R$ is defined as it was in the last section. Now, ultimately we
want to construct $SO(N)$ invariant operators, which is achieved by contracting the indices
in $A$ in pairs\footnote{If we contract indices with any tensor that is $SO(N)$ invariant,
we will produce a gauge invariant operator. Contracting indices in pairs corresponds to
contracting with $\delta^{ij}$s; $\delta^{ij}$ is indeed invariant under $SO(N)$. There is
a second $SO(N)$ tensor we could use: $\epsilon^{i_1\, i_2 \cdots i_n}$. We will return to this
possibility.}. Recall the rule for generating a tensor in a certain representation $R$.
Each index of a tensor is assigned to a box in Young diagram $R$. Indices in the same row
are then subject to symmetrization. Indices in the same column are then antisymmetrized.
Thus, if we want to contract all the indices, the only Young diagrams which give a non-zero
answer have an even number of boxes in each row. Further, the tensor transforming in representation
$R$ is not obtained by taking a trace of $n$ arbitrary antisymmetric tensors - we are tensoring
$n$ copies of a single antisymmetric tensor $Z^{a_{1,1}a_{1,2}}$. As a consequence, in the end,
the only Young diagrams which give a non-zero operator have an even number of boxes in each column.
Thus, for an $R$ that does give rise to a gauge invariant operator we need an even number of boxes 
in each row and an even number of boxes in each column. From the results above we see that there
are no gauge invariant operators for $n=1,3$. 
A little reflection shows that these are the Young diagrams which contribute: For $n=2$ we have
a single gauge invariant operator corresponding to
\bea
  \yng(2,2)
\eea
For $n=4$ we have two gauge invariant operators corresponding to
\bea
  \yng(4,4)\qquad \yng(2,2,2,2)
\eea
For $n=6$ we have three gauge invariant operators
\bea
  \yng(6,6)\qquad \yng(4,4,2,2)\qquad \yng(2,2,2,2,2,2)
\eea
No Young diagrams contribute for $n$ odd, i.e. there are no gauge invariant operators for odd $n$.

Lets make some preliminary comments related to the counting of gauge invariant operators.
Since $Z^{ab}=-Z^{ba}$ we know that at $n=1$, ${\rm Tr}(Z)=0$ implying there are no gauge
invariant operators. At $n=3$ one might hope that ${\rm Tr}(Z^3)$ gives a gauge invariant
operator; it doesn't, since
\bea
  {\rm Tr}(Z^3) &=&  Z^{ab}Z^{bc}Z^{ca}\cr
                &=& -Z^{ba}Z^{cb}Z^{ac}\cr
                &=& -Z^{cb}Z^{ba}Z^{ac}\cr
                &=& - {\rm Tr}(Z^3)
\eea
Clearly then, with an obvious extension of this argument, at all odd $n$ we don't have any gauge 
invariant operators. At $n=2$ we have ${\rm Tr}(Z^2)$, while at $n=4$ we have ${\rm Tr}(Z^4)$ and 
${\rm Tr}(Z^2)^2$. It is not difficult to put these operators in to a one-to-one correspondence 
with the Young diagrams that we identified above. There are four indices associated
to two $Z$'s - hence $R$ should have 4 blocks. This suggests
\bea
  {\rm Tr} (Z^2)\leftrightarrow \yng(2,2)
\eea
In general associate any single trace operator of $2n$ fields with a Young diagram $R$ that
has $2n$ boxes in the first row and $2n$ boxes in the second row. Here are a few examples:
\bea
  {\rm Tr}(Z^4)\leftrightarrow \yng(4,4)
\eea
\bea
  {\rm Tr}(Z^6)\leftrightarrow \yng(6,6)
\eea
If we have multitrace operators, stack the single trace diagram to obtain a valid diagram after
stacking. For example
\bea
  {\rm Tr}(Z^4){\rm Tr}(Z^2)\leftrightarrow \yng(4,4,2,2)
\eea
Looking back at the diagrams we found for $n=6$ we see that these correspond to the operators
${\rm Tr}(Z^6)$, ${\rm Tr}(Z^4){\rm Tr}(Z^2)$ and ${\rm Tr}(Z^2)^3$. 
This correspondence is strong evidence that the counting of Young diagrams with an even number of boxes in
each row and in each column matches the counting of gauge invariant operators for $SO(N)$ gauge theory.
We know by construction that the operators constructed using distinct Young diagrams $R$ are orthogonal
with respect to the free $SO(N)$ gauge theory two point function. 

To complete the construction of our gauge invariant operators, we now need to specify how indices are to be contracted in pairs. 
A new feature, not present in our toy model discussion, is that there are a number of inequivalent ways in which this can be done.
A particularly useful contraction for the purpose of constructing a basis is given by separating the indices into collections of four
adjacent indices and then contracting the outer two indices and the inner two indices in each collection.
With this contraction, our gauge invariant operators are
\bea
  O_R (Z) ={1\over (2n)!}\sum_{\sigma\in S_{2n}}\chi_R(\sigma )
           \sigma^{ i_1 i_2 i_2 i_1\cdots i_{n-1 }i_n i_n i_{n-1}}_{j_1 j_2 \cdots j_{2n-1} j_{2n}}
            Z^{j_1 j_2}Z^{j_3 j_4}\cdots Z^{j_{2n-1} j_{2n}}
\label{opone}
\eea
\bea
  \bar{O}_R (Z) ={1\over (2n)!}\sum_{\sigma\in S_{2n}}\chi_R(\sigma )\sigma_{ i_1 i_2 i_2 i_1 \cdots i_{n-1}i_n i_n i_{n-1}}^{j_1 j_2 \cdots j_{2n-1} j_{2n}}
            Z_{j_1 j_2}Z_{j_3 j_4}\cdots Z_{j_{2n-1} j_{2n}}
\label{optwo}
\eea
and $n$ must be even. Thus, $2n$ is divisible by $4$. The only reps $R$ which are allowed are built from the ``basic
block'' ${\tiny \yng(2,2)}$. Thus, the number of gauge invariant operators built using $n$ fields is equal to the number
of partitions of ${n\over 2}$. For $n=6$ we are talking about partitions of 3 and the association between partitions of
${n\over 2}$ and Young diagrams $R$ goes as follows
\bea
  [3]\leftrightarrow \yng(3) \leftrightarrow \yng(6,6)
  \label{corrone}
\eea
\bea
  [2,1]\leftrightarrow \yng(2,1) \leftrightarrow \yng(4,4,2,2)
  \label{corrtwo}
\eea
\bea
  [1,1,1]\leftrightarrow \yng(1,1,1) \leftrightarrow \yng(2,2,2,2,2,2)
  \label{corrthree}
\eea
In what follows we denote the partition of ${n\over 2}$ corresponding to Young diagram $R$ by $R/4$.
Thus, $R$ is a Young diagram with $2n$ boxes and $R/4$ is a Young diagram with ${n\over 2}$ boxes, related
as spelt out in the examples (\ref{corrone}), (\ref{corrtwo}) and (\ref{corrthree}) above.

We will now consider the two point functions of our operators.
Towards this end, consider the Wick contractions.
Our goal is to show that the Wick contractions are a sum over the wreath product $S_n [S_2]$, and further
that they again define a projection operator, projecting to a specific representation of $S_n [S_2]$.

The Wick contractions take the form
\bea
\langle T^A T_B\rangle = \sum_{\sigma\in S_n}\prod_{i=1}^n\sum_{\gamma_i\in S_2}{\rm sgn}(\gamma_i)
                         \delta^{a_{i1}}_{b_{\sigma(i)}\gamma_i (1)}\delta^{a_{i2}}_{b_{\sigma(i)}\gamma_i (2)}
\label{WickWreath}
\eea
This is a projector onto a specific representation of the wreath product $S_n[S_2]$.
To see which representation, note that a projector onto a representation of the wreath 
product can be written as
\bea
  \cP_{[A]}={1\over n! 2^n}\sum_{\rho\in S_n[S_2]}\chi_{[A]}(\rho ) \rho
\eea
We can give an element $\rho\in S_n[S_2]$ in terms of an element $\sigma\in S_n$ and $n$ elements
$\gamma_i\in S_2$ as we have done in (\ref{WickWreath}). Call this element 
$(\g_1,\g_2,...,\g_n;\sigma)$. We see that
\bea
  \chi_{[A]}\left((\g_1,\g_2,...,\g_n;\sigma)\right)=\prod_{i=1}^n {\rm sgn} (\g_i)
\eea
The representation with these characters can be obtained by tensoring $n$ copies of the antisymmetric 
representation of $S_2$ (i.e. {\tiny $\yng(1,1)$}) for each of the $\g_i$ with the identity representation for $\s$. 
The resulting representation is 1 dimensional and hence it is clearly irreducible.
The group elements in this representation are
\bea
  \Gamma^{[A]} \left( (\g_1,\g_2,...,\g_n;\sigma)\right) = \prod_{i=1}^n {\rm sgn} (\g_i)
  \label{wreathrep}
\eea
or equivalently
\bea
  \Gamma^{[A]} \left( \rho\right) =  {\rm sgn} (\rho)\qquad\rho\in S_n[S_2]
  \label{wreathrep}
\eea
Since this representation is 1 dimensional, the characters of this irreducible 
representation are just the group elements themselves. 

Since the Wick contractions define a projection operator which projects onto (\ref{wreathrep}),
we know that we can define an orthogonal gauge invariant operator for every copy of $[A]$ that is
subduced by a Young diagram $R$ with an even number of boxes in each row and column.
The number of times that $[A]$ is subduced by $R$ is given by
\bea
   p_R={1\over |S_n[S_2]|}\sum_{\rho\in S_n[S_2]}\chi_{[A]}(\rho)\chi_R(\rho)
\eea
We have evaluated this sum for all possible $R$ at $n=2,4$ and have verified that for each $R$ we obtain $p_R=1$.
This implies that there is only a single gauge invariant operator that can be defined for each $R$.
In fact, from existing group theory results we know that the irrep $[A]$ of $S_n[S_2]$ is always subduced once\cite{little}. 
To see this, note that Littlewood has proved that if we use $[A]$ to induce a representation of $S_{2n}$ this representation 
is reducible and decomposes into a multiplicity free sum\cite{little}.
By Frobenius reciprocity we know therefore that the appearance of $[A]$ in the restriction to $S_n[S_2]$ will also be multiplicity free.
This observation completes the logic relating the number of gauge invariant operators for $SO(N)$ gauge theory
and the number of Young diagrams with an even number of boxes in each row and in each column.

With the identification of the sum of Wick contractions as a permutation operator, we are now ready to tackle the computation of the
two point functions of our operators (to move to the last line below we have used the product of projection operators $\cP_R$ and $\cP_S$
which is reviewed in Appendix \ref{projprod})
\bea
   \langle O_R(Z)\bar{O}_S(Z)\rangle &=&
   \langle (\cP_R)^{i_1 i_2 i_2 i_1\cdots i_{n-1} i_n i_n i_{n-1}}_B T^B (\cP_S)^C_{j_1 j_2 j_2 j_1\cdots j_{n-1} j_n j_n j_{n-1}}T_C\rangle\cr
&=& n!2^n (\cP_R\cP_{[A]}\cP_S)^{i_1 i_2 i_2 i_1\cdots i_{n-1} i_n i_n i_{n-1}}_{j_1 j_2 j_2 j_1\cdots j_{n-1} j_n j_n j_{n-1}}\cr
&=& n!2^n (\cP_R \cP_S \cP_{[A]})^{i_1 i_2 i_2 i_1\cdots i_{n-1} i_n i_n i_{n-1}}_{j_1 j_2 j_2 j_1\cdots j_{n-1} j_n j_n j_{n-1}}\cr
&=& {\delta_{RS}\over d_R} n!2^n (\cP_R\cP_{[A]})^{i_1 i_2 i_2 i_1\cdots i_{n-1} i_n i_n i_{n-1}}_{j_1 j_2 j_2 j_1\cdots j_{n-1} j_n j_n j_{n-1}}
\label{twopointcomputation}
\eea
The product of the projection operators appearing in this last line is 
\bea
\cP_R\cP_{[A]} 
&=& {1\over 2^n n! (2n)!}\sum_{\rho\in S_{2n}}\sum_{\sigma\in S_n\big[ S_2\big]}\chi_R(\rho)\chi_{[A]}(\sigma)\rho\sigma\cr
&=& {1\over 2^n n! (2n)!}\sum_{\psi\in S_{2n}}\sum_{\sigma\in S_n\big[ S_2\big]}\chi_R(\psi\sigma^{-1})\chi_{[A]}(\sigma)\psi\cr
&=&{1\over (2n)!}\sum_{\psi\in S_{2n}} {\rm Tr}(P_{[A]}\Gamma^R(\psi))\psi
\label{nonstdprojprod}
\eea
Thus, we now find
\bea
   \langle O_R(Z)\bar{O}_S(Z)\rangle
  ={\delta_{RS}\, n!2^n \over (2n)!\, d_R} \sum_{\psi\in S_{2n}} {\rm Tr}(P_{[A]}\Gamma^R(\psi))
(\psi)^{i_1 i_2 i_2 i_1\cdots i_{n-1} i_n i_n i_{n-1}}_{j_1 j_2 j_2 j_1\cdots j_{n-1} j_n j_n j_{n-1}}
\label{twpnt}
\eea
Before we proceed further recall that $P_{[A]}$ projects onto the $[A]$ irrep of $S_n[S_2]$,
where $S_n[S_2]$ is the stabilizer of the cycle $(1,2)(3,4)(5,6)\cdots (2n-1,2n)$. We now want to introduce
a second $S_n[S_2]$ which has a different embedding into $S_{2n}$: the new $S_n[S_2]$ is the stabilizer of
$(2,3)(4,1)(6,7)(8,5)\cdots (2n-2,2n-1)(2n,2n-3)$. 
Then, introduce the Brauer algebra ${\cal B}_n$ of size $n$, which is the set of pairings of $\{1,2,...,2n\}$. 
As an example, ${\cal B}_2$ has 3 elements given by $(1,2)(3,4)$, $(1,3)(2,4)$ and $(1,4)(2,3)$.
The importance of the Brauer algebra follows from the fact that
${\cal B}_n$ is isomorphic to $S_{2n}/S_n[S_2]$. Thus, we can write
\bea
\langle O_R(Z)\bar{O}_S(Z)\rangle
&=&{\delta_{RS}\, n!2^n \over (2n)!\, d_R} \sum_{\psi_1\in {\cal B}_n}\sum_{\psi_2\in S_n[S_2]} {\rm Tr}(P_{[A]}\Gamma^R(\psi_1\psi_2))
(\psi_1\psi_2)^{i_1 i_2 i_2 i_1\cdots i_{n-1} i_n i_n i_{n-1}}_{j_1 j_2 j_2 j_1\cdots j_{n-1} j_n j_n j_{n-1}}\cr
&=&{\delta_{RS}\, n!2^n \over (2n)!\, d_R} \sum_{\psi_1\in {\cal B}_n}\sum_{\psi_2\in S_n[S_2]} {\rm Tr}(P_{[A]}\Gamma^R(\psi_1\psi_2))
(\psi_1)^{i_1 i_2 i_2 i_1\cdots i_{n-1} i_n i_n i_{n-1}}_{j_1 j_2 j_2 j_1\cdots j_{n-1} j_n j_n j_{n-1}}\cr
&=&{\delta_{RS}\, (n!2^n)^2 \over (2n)!\, d_R} \sum_{\psi_1\in {\cal B}_n} {\rm Tr}(P_{[A]}\Gamma^R(\psi_1)\hat{P}_{[S]})
(\psi_1)^{i_1 i_2 i_2 i_1\cdots i_{n-1} i_n i_n i_{n-1}}_{j_1 j_2 j_2 j_1\cdots j_{n-1} j_n j_n j_{n-1}}
\eea
In the second last line we have recognized the fact that $S_n[S_2]$ acts trivially on the indices $j_1 j_2 j_2 j_1\cdots j_{n-1} j_n j_n j_{n-1}$.
In the last line above we have introduced the projector
\bea
\hat{P}_{[S]}={1\over 2^n n!}\sum_{\psi_2\in S_n[S_2]}\Gamma^R(\psi_2)
\eea
which projects onto the irreducible representation $[S]$ of $S_n [S_2]$; the irrep $[S]$ represents every element of $S_n[S_2]$ by 1. 
We have written the projector $P_{[A]}$ without a hat and the projector $\hat{P}_{[S]}$ with a hat to remind us that they
project to representations that belong to $S_n[S_2]$ subgroups with different embeddings\footnote{$P_{[A]}$ is defined by
using the $S_n[S_2]$ subgroup that stabilizes $(1,2)(3,4)\cdots (2n-1,2n)$; $\hat{P}_{[S]}$ is defined using the subgroup
that stabilizes $(2,3)(4,1)\cdots (2n-2,2n-1)(2n,2n-3)$. The Brauer algebra is a set of coset representatives for $S_{2n}/S_n[S_2]$
where the embedding of this $S_n[S_2]$ is the same as the embedding used to construct $\hat{P}_{[S]}$.}. We will now focus on the sum
over ${\cal B}_n$
\bea
  S=\sum_{\psi\in {\cal B}_n}{\rm Tr}(P_{[A]}\Gamma^R (\psi)\hat{P}_{[S]})
    (\psi)^{i_1 i_2 i_2 i_1\cdots i_{n-1} i_n i_n i_{n-1}}_{j_1 j_2 j_2 j_1\cdots j_{n-1} j_n j_n j_{n-1}}
\eea
By permuting slots we can rewrite this as
\bea
  S=\sum_{\psi\in {\cal B}_n}{\rm Tr}(P_{[A]}\Gamma^R (\psi)\hat{P}_{[S]})(\psi)^{i_1 i_1 i_2 i_2\cdots i_n i_n}_{j_1 j_1 j_2 j_2 \cdots j_n j_n}
\eea
To evaluate this sum we can now proceed using the logic of Appendix \ref{detailed}.
The basic idea is that the choice of elements of ${\cal B}_n$ is not unique; each element of
${\cal B}_n$ corresponds to a coset of $S_{2n}$ with respect to the subgroup $S_n[S_2]$ and 
we have freedom in choosing the coset representative.
It is possible to choose a set of representatives such that
\bea
  S={\rm Tr}_{R}(P_{[A]}\prod_{i=1}^{n-1}[N+J_{2i-1}]\hat{P}_{[S]})
  \label{middlestep}
\eea
with $J_{2n-1}$ the Jucys-Murphy element in irrep $R$ given by
\bea
  J_{2n-1}=\sum_{i=1}^{2n-2} \Gamma^R ((2n-1,i))
\eea
We have explicitly indicated that the trace in (\ref{middlestep}) runs over the carrier space of $R$.
Now, $P_{[A]}$ forces antisymmetry between boxes $1$ and $2$, $3$ and $4$,..., $2n-1$ and $2n$. 
$\hat{P}_{[S]}$ forces symmetry between boxes $2$ and $3$, $4$ and $1$,..., $2n-2$ and $2n-1$, $2n$ and $2n-3$.
For $n=4$ the pair select the following state
\bea
\young(42,31)
\eea 
For $n=8$ and $R={\tiny \yng(4,4)}$ for example, the following state is selected by $P_{[A]}$
\bea
\young(8642,7531)
\eea
In general, there is more than one state selected by $\hat{P}_{[S]}$, but in the end all states that contribute 
have the odd integers in even rows, where the topmost row is labeled 1. 
Thus, for example, the odd integers would appear in the rows with a $*$ below
\bea
  \young(\,\,\,\,,****,\,\,,**)
\eea
Lets call these the ``odd boxes''. We now find that
\bea
  S={\rm Tr}_{R}(P_{[A]}\hat{P}_{[S]})\prod_{i\in {\rm odd\,\, boxes\,\, in\,\,}R}c_i
\eea
with $c_i$ the factor of the box $i$. 
Recall that a box in row $i$ and column $j$ has factor $N+j-i$.
Thus,
\bea
\langle O_R(Z)\bar{O}_S(Z)\rangle
={\delta_{RS}\, (n!2^n)^2 \over (2n)!\, d_R} {\rm Tr}_{R}(P_{[A]}\hat{P}_{[S]})\prod_{i\in {\rm odd\,\, boxes\,\, in\,\,}R}c_i
\eea
To compute the remaining trace, notice that by employing the permutation
\bea
  \rho = (1,2,3,4)(5,6,7,8)\cdots (2n-3,2n-2,2n-1,2n)
\eea
we can write
\bea
  {\rm Tr}_{R}(P_{[A]}\hat{P}_{[S]})={\rm Tr}_{R}(P_{[A]}\rho P_{[S]}\rho^{-1})
\eea
Since both $[S]$ and $[A]$ are one dimensional representations, we know that the projectors $P_{[A]}$ and $P_{[S]}$ can be written
as the outer product of a single vector
\bea
 P_{[A]}=|[A]\rangle\langle [A]|\qquad P_{[S]} =|[S]\rangle\langle[S]|
\eea
Thus, assuming without loss of generality that $R$ is an orthogonal representation,
the trace we are interested in can be written as the square of a single matrix element
\bea
  {\rm Tr}_{R}(P_{[A]}\hat{P}_{[S]}) = ( \langle [A] | \Gamma^R(\rho) | [S] \rangle )^2
\eea
Remarkably, precisely this matrix element has been studied in \cite{ivanov}.
We review the relevant results of \cite{ivanov} in Appendix \ref{ivanovstuff}.
From those results we now find
\bea
\langle O_R(Z)\bar{O}_S(Z)\rangle
=\delta_{RS}\, 2^{n}\left( {d_{R/4} \over  d_R}\right)^2\,\prod_{i\in {\rm odd\,\, boxes\,\, in\,\,}R}c_i
\label{mainresult}
\eea
which completes the evaluation of the two point function.

Although they do not form a basis, we will also make use of the operators
\bea
  W_R (Z) ={1\over (2n)!}\sum_{\sigma\in S_{2n}}\chi_R(\sigma )
           \sigma^{ i_1 i_2 i_2 i_3 i_3\cdots i_n i_n i_{1}}_{j_1 j_2 j_3 j_4\cdots j_{2n-1} j_{2n}}
            Z^{j_1 j_2}Z^{j_3 j_4}\cdots Z^{j_{2n-1} j_{2n}}
\label{oldopone}
\eea
\bea
 \bar{W}_R (Z) ={1\over (2n)!}\sum_{\sigma\in S_{2n}}\chi_R(\sigma)\sigma_{i_1 i_2 i_2 i_3 i_3\cdots i_n i_n i_1}^{j_1 j_2 j_3 j_4 \cdots j_{2n-1}j_{2n}}
            Z_{j_1 j_2}Z_{j_3 j_4}\cdots Z_{j_{2n-1} j_{2n}}
\label{oldoptwo}
\eea
These operators do not form a basis because they are only non-zero when $R/4$ is a hook. Indeed, the two point functions are
computed using (C.6). This involved a character of an ${n\over 2}$ cycle which vanishes unless $R/4$ is a hook.
Again using the results of \cite{ivanov} we find, when $R/4$ is a hook
\bea
\langle W_R(Z)\bar{W}_S(Z)\rangle
=\delta_{RS}\, \left({2 \over d_R}\right)^2 \,\prod_{i\in {\rm odd\,\, boxes\,\, in\,\,}R}c_i
\label{mainresult}
\eea
and when $R/4$ is not a hook
\bea
\langle W_R(Z)\bar{W}_S(Z)\rangle
=0
\label{secondmainresult}
\eea

\section{Back to the trace basis}

It is well established in $U(N)$ gauge theory, that correlations functions of operators in the trace basis can be computed using the 
correlation functions of the Schur polynomials.
This allows computations which are exact to all order in $1/N$\cite{Corley:2002mj,Caputa:2012dg}. 
In this section we will explain how to use the correlation functions of the operators we have introduced, for the same purpose.
This again allows computations which are exact to all orders in $1/N$, in the $SO(N)$ gauge theory.

Our approach is a very natural extension of the Schur polynomial technology and it again makes heavy use of the character
orthogonality relation which reads
\bea
  \sum_R \chi_R (\sigma)\chi_R (\rho) = {|{\cal G}|\over |[\sigma]|}\delta_{[\sigma][\rho]}
\eea
Here the sum runs over a complete set of irreducible representations of group ${\cal G}$,
$|{\cal G}|$ is the order of group ${\cal G}$, $|[\sigma]|$ is the number of elements in conjugacy
class $[\sigma]$ and $\delta_{[\sigma][\rho]}$ is 1 if $\sigma$ and $\rho$ belong to the same
conjugacy class and is zero otherwise. Using this result we find
\bea
    \sum_R \chi_R (\psi)  O_R (Z) 
     &=& {1\over (2n)!}\sum_{\sigma\in S_{2n}}\sum_R \chi_R (\psi)\chi_R(\sigma )
            \sigma^{ i_1 i_2 i_2 i_1\cdots i_{n-1} i_n i_n i_{n-1}}_{j_1 j_2 \cdots j_{2n-1} j_{2n-2}}
            Z^{j_1 j_2}Z^{j_3 j_4}\cdots Z^{j_{2n-1} j_{2n}}\cr
     &=& {1\over |[\psi]|}\sum_{\sigma\in [\psi]}
            \sigma^{ i_1 i_2 i_2 i_1 \cdots i_{n-1} i_n i_n i_{n-1}}_{j_1 j_2 \cdots j_{2n-1} j_{2n-2}}
            Z^{j_1 j_2}Z^{j_3 j_4}\cdots Z^{j_{2n-1} j_{2n}}
\eea
If we take $\psi = 1$ the identity element, then $|[\psi]|=1$, $\chi_R (\psi)=d_R$ so that we easily find
\bea
  {\rm Tr}(Z^2)^{n\over 2}=    \sum_{R/4\vdash {n\over 2}} d_R O_R (Z) 
\eea
Denoting ${\rm Tr}(Z^2)=Z^{ij}Z^{ji}$ and ${\rm Tr}(\bar{Z}^2)=Z_{ij}Z_{ji}$ we easily find
\bea
  \langle {\rm Tr}(Z^2)^{n\over 2}{\rm Tr}(\bar{Z}^2)^{n\over 2}\rangle &=& \sum_{R/4,S/4\,\vdash {n\over 2}}
           d_R d_S\langle O_R(Z)\bar{O}_S(Z)\rangle\cr
           &=&2^n\sum_{R/4\vdash {n\over 2}}(d_{R/4})^2\,\prod_{i\in {\rm odd\,\, boxes\,\, in\,\,}R}c_i
\eea
At large $N$ with $n$ fixed to be order $1$ we have
\bea
   \prod_{i\in {\rm odd\,\, boxes\,\, in\,\,}R}c_i = N^n\left(1+O\left({1\over N}\right)\right)
\eea
so that
\bea
  \langle {\rm Tr}(Z^2)^{n\over 2}{\rm Tr}(\bar{Z}^2)^{n\over 2}\rangle 
           &=& 2^n N^n \sum_{R/4 \vdash {n\over 2}}(d_{R/4})^2\left(1+O\left({1\over N}\right)\right)\cr
           &=& 2^n N^n \left({n\over 2}\right)!\left(1+O\left({1\over N}\right)\right)
\eea
This is the correct large $N$ result for these correlators.

A very similar argument gives
\bea
    \sum_R \chi_R (\psi)  W_R (Z) 
     &=& {1\over (2n)!}\sum_{\sigma\in S_{2n}}\sum_R \chi_R (\psi)\chi_R(\sigma )
            \sigma^{ i_1 i_2 i_2 \cdots i_n i_n i_1}_{j_1 j_2 \cdots j_{2n-1} j_{2n-2}}
            Z^{j_1 j_2}Z^{j_3 j_4}\cdots Z^{j_{2n-1} j_{2n}}\cr
     &=& {1\over |[\psi]|}\sum_{\sigma\in [\psi]}
            \sigma^{ i_1 i_2 i_2 \cdots i_n i_n i_1}_{j_1 j_2 \cdots j_{2n-1} j_{2n-2}}
            Z^{j_1 j_2}Z^{j_3 j_4}\cdots Z^{j_{2n-1} j_{2n}}
\eea
Again taking $\psi = 1$, we have $|[\psi]|=1$ and $\chi_R (\psi)=d_R$ so that we easily find
\bea
  {\rm Tr}(Z^n)=    \sum_R d_R  W_R (Z) 
\eea
Following the same steps as above, we now find
\bea
  \langle {\rm Tr}(Z^n){\rm Tr}(\bar{Z}^n)\rangle 
      = \sum_{R/4{\rm \,\, is \,\, a\,\, hook}}4\prod_{i\in {\rm odd\,\, boxes\,\, in\,\,}R}c_i
\eea
The sum over $R$ can be parametrized in the following way. Denoting $n=2J$, $R/4$ consists of $J$ hook diagrams 
with $J$ boxes each. They can be labelled by the number of boxes in the first column $k\in\left\{1,...,J\right\}$. 
Then the product over weights of odd boxes in $R$ reads
\begin{eqnarray}
\prod_{i} c_i=\prod^{2(J-k)}_{i=1}(N+i)\prod^{2k}_{j=1}(N-j+1)=\frac{\G(N+2(J-k)+1)}{\G(N-2k+1)}.
\end{eqnarray}
Adding all together
\begin{eqnarray}
\langle Tr(Z^{2J})Tr(\bar{Z}^{2J})\rangle&=&4\sum^{J}_{k=1}\frac{\G(N+2(J-k)+1)}{\G(N-2k+1)}\nonumber\\
&=&4J\,N^{2J}\left(1-\frac{J}{N}+\frac{J(J-1)(4J^2-1)}{6N^2}-O\left(\frac{1}{N^3}\right)\right).\nonumber\\
\end{eqnarray}
Note that the large $N$ result is the same as in the $U(N)$ gauge group\footnote{Relative factor of 2 comes from 
our normalization of the $SO(N)$ generators} and 
we also have the breakdown of the color expansion for $J\sim \sqrt{N}$. 
The novelty is the 1/N correction that corresponds to diagrams with a cross-cap.

\section{Further gauge invariant operators}

There are two $O(N)$ invariant tensors: $\delta^{ij}$ and $\epsilon^{i_1 i_2 \cdots i_N}$.
We assume that $N$ is even. Whenever $(T_R)^A$ in (\ref{basicformula}) has more than $N$
indices we can use $\epsilon^{i_1 i_2 \cdots i_N}$ together with $\delta^{ij}$s to produce 
a gauge invariant operator. Consider the case of exactly $N$ indices - in this case we can 
construct the gauge invariant operator
\bea
  Q_R (Z) =\epsilon_{ i_1 i_2 i_3 \cdots i_N}
           {1\over N!}\sum_{\sigma\in S_{N}}\chi_R(\sigma )\sigma^{ i_1 i_2 i_3 \cdots i_{N-1} i_N}_{j_1 j_2 j_3\cdots j_{N-1} j_{N}}
            Z^{j_1 j_2}Z^{j_3 j_4}\cdots Z^{j_{N-1} j_{N}}
\eea
Now, we know that, to get a non-zero result we need to choose an $R$ that leads to a tensor antisymmetric in all $N$ indices.
This forces us to take $R=1^N$. In this case
\bea
{1\over N!}\sum_{\sigma\in S_{N}}\chi_{1^N}(\sigma )\sigma^{ i_1 i_2 i_3 \cdots i_{N-1} i_N}_{j_1 j_2 j_3 \cdots j_{N-1} j_{N}}
={1\over N!}\epsilon^{i_1 i_2 i_3 \cdots i_N}\epsilon_{j_1 j_2 j_3 \cdots j_N}
\eea
Thus, it is now simple to see that
\bea
  Q_{1^N} (Z) =\epsilon_{j_1\, j_2\, j_3\, j_4 \, \cdots \, j_{N-1}\, j_{N}} Z^{j_1 j_2}Z^{j_3 j_4}\cdots Z^{j_{N-1} j_{N}}
\eea
which is (up to normalization) the Pfaffian. Notice that $R=1^N$ is distinct 
from the representations we use in constructing (\ref{opone}), so that
once again the Pfaffian is orthogonal to every other operator we have constructed. The computation of the two point function
is done exactly as we did it above. The result is (see also the derivation in \cite{Aharony:2002nd})
\bea
\langle Q_{1^N}(Z)\bar{Q}_{1^N}(Z)\rangle &=& {\left({N\over 2}\right)! 2^{N\over 2}\over N!}
                            \sum_{\psi\in S_N}{\rm Tr}(P_{[A]}\Gamma^{1^N}(\psi))
                              \psi^{ i_1 i_2 i_3 \cdots i_{N-1} i_N}_{j_1 j_2 j_3 \cdots j_{N-1} j_{N}}
\epsilon_{i_1 i_2 i_3 \cdots i_N}\epsilon^{j_1 j_2 j_3 \cdots j_N}\cr
&=&{\left({N\over 2}\right)! 2^{N\over 2}\over N!}
                            \sum_{\psi\in S_N}{\rm Tr}(\Gamma^{1^N}(\psi))
                              \psi^{ i_1 i_2 i_3 \cdots i_{N-1} i_N}_{j_1 j_2 j_3 \cdots j_{N-1} j_{N}}
\epsilon_{i_1 i_2 i_3 \cdots i_N}\epsilon^{j_1 j_2 j_3 \cdots j_N}\cr
&=&\left({N\over 2}\right)! 2^{N\over 2} N!
\eea
Lets do a few checks. For $N=2$,
\bea
Q_{1^2}= Z^{12}-Z^{21}=2Z^{12}
\eea
Thus,
\bea
   \langle Q_{1^2}\bar{Q}_{1^2}\rangle = 4 \langle Z^{12}Z_{12}\rangle = 4 = \left({2\over 2}\right)! 2^{2\over 2} 2!
\eea
which matches our result.
For $N=4$ we have
\bea
Q_{1^4}= 8 (Z^{12}Z^{34}+Z^{13}Z^{42}+Z^{14}Z^{23})
\eea
Thus,
\bea
   \langle Q_{1^4}\bar{Q}_{1^4}\rangle =  64 \langle (Z^{12}Z^{34}+Z^{13}Z^{42}+Z^{14}Z^{23})(Z_{12}Z_{34}+Z_{13}Z_{42}+Z_{14}Z_{23})\rangle \cr
                           = 192= \left({4\over 2}\right)! 2^{4\over 2} 4!
\eea
which again matches our result.

Now, consider an operator built using $N/2 +2$ fields and use a combination of the $\epsilon^{i_1 i_2\cdots i_N}$ and $\delta^{ij}$s
to produce the gauge invariant operator
\bea
Q_R (Z)={\epsilon_{i_1 i_2\cdots i_N} \over (N+4)!}\sum_{\sigma\in S_{N+4}}\chi_R(\sigma)
        \sigma^{i_1 i_2\cdots i_{N-1}i_N i_{N+1} i_{N+2} i_{N+2} i_{N+1}}_{j_1 j_2\cdots j_N j_{N+1} j_{N+2} j_{N+3} j_{N+4}}
         Z^{j_1 j_2}Z^{j_3 j_4}\cdots Z^{j_{N+1} j_{N+2}}Z^{j_{N+3} j_{N+4}}\cr
\eea
Arguing as we did above, to get a non-zero result we need to consider a Young diagram $R\vdash N+4$ that has parts $(3^2,1^{N-2})$.
Since $R$ is distinct from all other representations we have used, operators constructed in this way will again be orthogonal to
operators constructed using only the $\delta^{ij}$s.
The two point function of this operator is
\bea
  \langle Q_R (Z)\bar{Q}_R(Z)\rangle = {\left({N+4\over 2}\right)! 2^{N+4\over 2}\over (N+4)!}\sum_{\psi\in S_{N+4}}
       {\rm Tr}(P_{[A]}\Gamma^R(\psi))
       \psi^{i_1 i_2\cdots i_{N-1}i_Ni_{N+1}i_{N+2}i_{N+2}i_{N+1}}_{j_1j_2\cdots j_{N-1}j_Nj_{N+1}j_{N+2}j_{N+2}j_{N+1}}
       \epsilon_{i_1 i_2\cdots i_N}\epsilon^{j_1 j_2\cdots j_N}\cr
\eea
The sum that needs to be performed here has a different structure to the sum appearing in (\ref{twpnt}) and it seems that the result
\cite{ivanov} does not help. We leave the evaluation of this sum as an interesting problem for the future.

Our results suggest that in the general case $R$ is a Young diagram with a single column of $N$ boxes adjoined to the right with a 
Young diagram which has both an even number of columns and an even number of rows. 
Operators that are constructed with an even number of $\epsilon_{i_1 i_2\cdots i_N}$s are equivalent to operators constructed using
only $\delta^{ij}$s.
Indeed, each epsilon forces $R$ to have a column with $N$ boxes so that in total $R$ has an even number of columns and rows.
The same argument implies that operators that are constructed with an odd number of $\epsilon_{i_1 i_2\cdots i_N}$s are equivalent 
to operators constructed using only one $\epsilon_{i_1 i_2\cdots i_N}$ and $\delta^{ij}$s.
Thus, we should restrict ourselves to using only $\delta^{ij}$s or using $\delta^{ij}$s and a single $\epsilon_{i_1 i_2\cdots i_N}$.

\section{Discussion and future directions}

We have developed techniques that allow the computation of correlation functions of large operators in $SO(N)$ gauge theory.
The structure of our solution is a genuinely novel extension of the corresponding solution for $U(N)$ gauge theory\cite{cjr}. 
We have started by constructing $SO(N)$ tensors that are orthogonal with respect to the free field theory inner product. 
We have then extracted gauge invariant operators by contracting all indices, using either $\epsilon^{i_1 i_2\cdots i_n}$s or $\delta^{ij}$s.
For the case that we use only $\delta^{ij}$s we have a rather complete answer in the form of an easy characterization of the
operators (in terms of Young diagrams that have both an even number of columns and an even number of rows) and a very
explicit formula for the two point function. 
When we use both $\epsilon^{i_1 i_2\cdots i_n}$s and $\delta^{ij}$s, our organization continues to diagonalize the two point
function.
We do not however have a nice simple characterization of our operators or a formula for their two point functions.
This should be tackled to complete the project we have started. 
We have used the results for the two point functions of our operators to compute correlators of traces, to all orders in $1/N$.

There are a number of interesting ways in which this work can be extended. 
First, by computing the free field partition function one would be able to count how many operators can be constructed using a single matrix. 
This number should be compared to the number operators we have constructed. 
For operators built using $n<{N\over 2}$ fields, this should be equal to the number of partitions of ${n\over 2}$.
For $n>{N\over 2}$ fields one can use the $\epsilon^{i_1 i_2\cdots i_n}$ tensor to construct gauge invariant operators.
Although we do not have a clear prediction in this case, our results suggest that the number of operators will be equal to the
number of partitions of ${n\over 2}$ plus the number of partitions of ${n\over 2}-{N\over 2}$. 
All partitions must be restricted to have at most ${N\over 2}$ parts.
This constraint implements the stringy exclusion principle.  
It would be very interesting to test this conjecture.

We have only considered two point functions. 
For the theory with gauge group $U(N)$ the technology for computing n-point correlators is rather 
well developed\cite{Corley:2002mj,Caputa:2012dg}.
A key ingredient is a product rule satisfied by Schur polynomials, known as the Littlewood-Richardson rule. 
Using this product rule, the computation of any n-point correlator is reduced to the computation of a two point correlator.
It is important to develop a product rule for the operators $O_R(Z)$ defined in (\ref{opone}) in order to achieve similar
results for the theory with gauge group $SO(N)$.

Another interesting extension would be to consider operators built using more than a single matrix.
This is needed before one can study the spectrum of anomalous dimensions of the theory, which would be another interesting question to pursue.

\noindent
{\it Acknowledgements:}
This work is based upon research supported by the South African Research Chairs
Initiative of the Department of Science and Technology and National Research Foundation.
Any opinion, findings and conclusions or recommendations expressed in this material
are those of the authors and therefore the NRF and DST do not accept any liability
with regard thereto.
RdMK would like to thank Collingwood College, Durham for their support.

\begin{appendix}

\section{Detailed evaluation of the vector model two point function}\label{detailed}

In this Appendix we will explain how to obtain (\ref{exactvector}). This amounts to evaluating
\bea
I =  \sum_{\sigma\in S_{2p}}\chi_{(2p)}(\sigma)\sigma^{i_1 i_1 \cdots i_p i_p}_{j_1 j_1 \cdots j_p j_p}
\eea
We will discuss this example in detail as the logic of this computation will appear again in our
discussion of the $SO(N)$ correlation functions.
We want to replace the sum over $S_{2p}$ by a sum over the subgroup $S_p[S_2]$ and its cosets.
Looking at the way that the indices on the permutation above are contracted, we choose the subgroup
$S_p[S_2]$ to be the stabilizer of $(1,2)(3,4)\cdots (2p-1,2p)$. 
The set of coset representatives is ${\cal B}_p$ the Brauer algebra of size $p$. 
The cardinality of this set is
\bea
  |{\cal B}_p|=(2p-1)!!
\eea
To construct the set ${\cal B}_p$
we need to choose $(2p-1)!!$ elements $\{ \psi_i\}$ out of $S_{2p}$ such that no two elements are related by
$\psi_i=\sigma\psi_j$ with $\sigma\in S_n[S_2]$. 
A particularly convenient choice is given by taking the terms obtained by expanding the product
\bea
(1,1)\times \sum_{i=1}^3 (i,3) \times \sum_{i=1}^5 (i,5) \times \cdots \times \sum_{i=1}^{2p-1} (i,2p-1) 
=\prod_{j=0}^{p-1}\sum_{i=1}^{2j+1}(i,2j+1)
\label{Bchoice}
\eea
This clearly gives the required number $(2p-1)!!$ of terms. 
It is further easy to verify that no two of these terms
can be related by $\psi_i=\sigma\psi_j$ with $\sigma\in S_n[S_2]$. 

{\vskip 0.2cm}

\noindent
{\bf Explicit example:} Consider $S_4$. The stabilizer of $(1,2)(3,4)$ is
\bea
S_2 [S_2]=\{ 1\,,\,(1,2)\,,\,(3,4)\,,\,(1,2)(3,4)\,,\,(1,3)(2,4)\,,\,(1,4,2,3)\,,\,(1,3,2,4)\,,\,(1,4)(2,3)\}
\nonumber
\eea
In this case, (\ref{Bchoice}) gives
\bea
  (1,1)[(1,3)+(2,3)+(3,3)]=1+(2,3)+(1,3)
  \nonumber
\eea
It is simple to verify that $S_4 = S_2[S_2]\,\cup \,(13)\times S_2[S_2]\, \cup \, (32)\times S_2[S_2]$.

{\vskip 0.2cm}

Thus, we now obtain
\bea
I =  \sum_{\sigma\in S_{2p}}\chi_{(2p)}(\sigma)\sigma^{i_1 i_1 \cdots i_p i_p}_{j_1 j_1 \cdots j_p j_p}\cr
  =  \sum_{\psi_1 \in {\cal B}_{p}}\sum_{\psi_2 \in S_{p}[S_2]}\chi_{(2p)}(\psi_1\psi_2)
          (\psi_1\psi_2)^{i_1 i_1 \cdots i_p i_p}_{j_1 j_1 \cdots j_p j_p}\cr
  =  \sum_{\psi_1 \in {\cal B}_{p}}\sum_{\psi_2 \in S_{p}[S_2]}\chi_{(2p)}(\psi_1\psi_2)
          (\psi_1)^{i_1 i_1 \cdots i_p i_p}_{j_1 j_1 \cdots j_p j_p}\cr
\cr
{\rm (to\, get\, this\, last\, line\, recognize\, that}\, \psi_2\, {\rm has\, a\, trivial\, action\, on}\, j\, {\rm indices)}\cr
\cr
  =  \sum_{\psi_1 \in {\cal B}_{p}}\sum_{\psi_2 \in S_{p}[S_2]}\chi_{(2p)}(\psi_1)
          (\psi_1)^{i_1 i_1 \cdots i_p i_p}_{j_1 j_1 \cdots j_p j_p}\cr
\cr
{\rm (to\, get\, this\, last\, line\, recognize\, that\, in\, irrep}\, (2p)\,{\rm all\, characters\, are\, 1)}\cr
\cr
=2^p p!\sum_{\psi_1 \in {\cal B}_{p}}\chi_{(2p)}(\psi_1)
          (\psi_1)^{i_1 i_1 \cdots i_p i_p}_{j_1 j_1 \cdots j_p j_p}
\eea
To proceed, we will now consider what the value of $(\psi_1)^{i_1 i_1 \cdots i_p i_p}_{j_1 j_1 \cdots j_p j_p}$ is.
The rule is rather simple. Imagine $p=4$. For $\psi_1=(2,3)(4,5)(6,7)$ we have
\bea
(\, (2,3)(4,5)(6,7)\,)^{i_1 i_1 i_2 i_2 i_3 i_3 i_4 i_4}_{j_1 j_1 j_2 j_2 j_3 j_3 j_4 j_4}
=\delta^{i_1}_{j_1}\delta^{i_1}_{j_2}\delta^{i_2}_{j_1}\delta^{i_2}_{j_3}
 \delta^{i_3}_{j_2}\delta^{i_3}_{j_4}\delta^{i_4}_{j_3}\delta^{i_4}_{j_4}
=N
\eea
This is shown diagramatically in the figure \ref{cycle1} below.
\begin{figure}[ht]%
\begin{center}
\includegraphics[width=6cm]{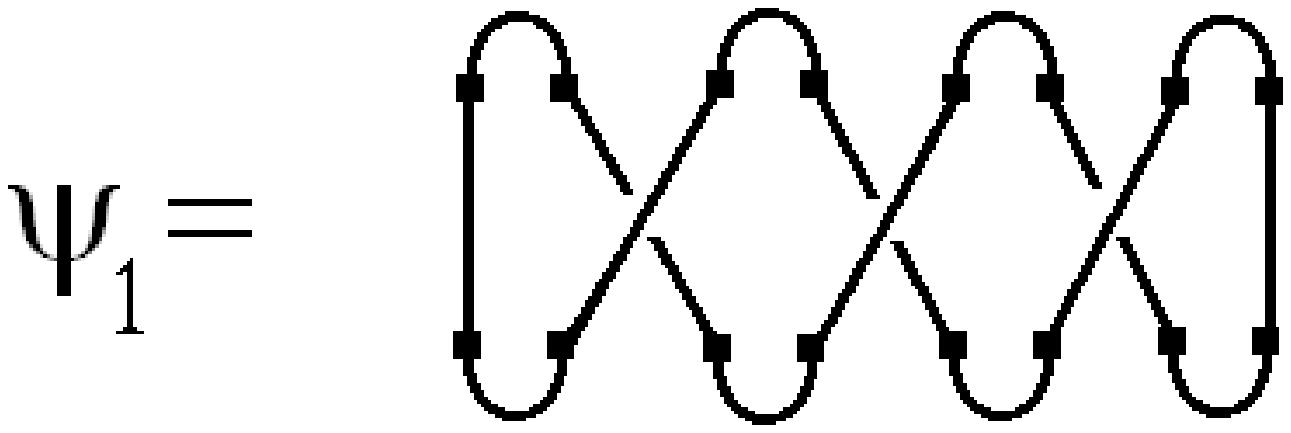}%
\caption{$(\psi_1)^{i_1i_1i_2i_2i_3i_3i_4i_4}_{j_1j_1j_2j_2j_3j_3j_4j_4}=N$ for $\psi_1 = (2,3)(4,5)(6,7) $.}%
\label{cycle1}%
\end{center}
\end{figure}

\begin{figure}[ht]%
\begin{center}
\includegraphics[width=6cm]{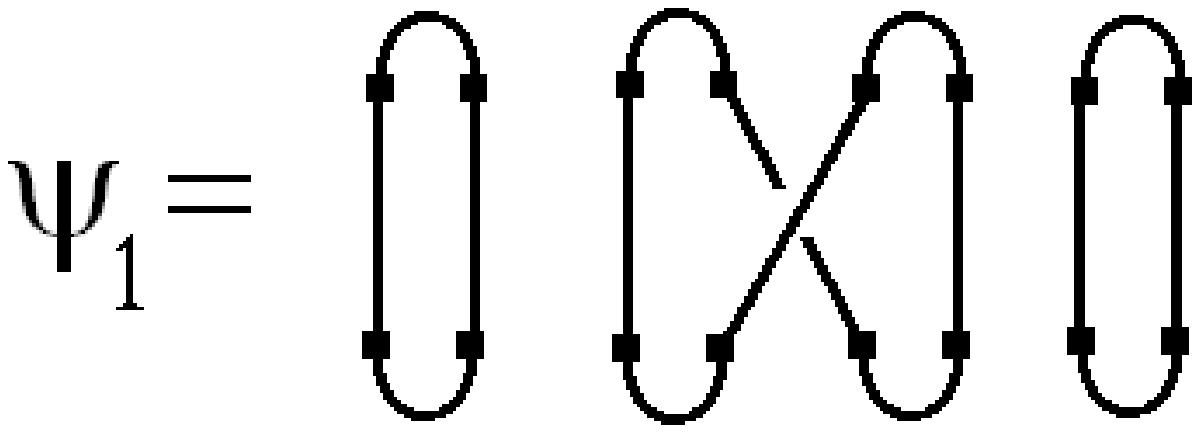}%
\caption{$(\psi_1)^{i_1i_1i_2i_2i_3i_3i_4i_4}_{j_1j_1j_2j_2j_3j_3j_4j_4}=N^3$ for $\psi_1 = (4,5) $.}%
\label{cycle2}%
\end{center}
\end{figure}

\begin{figure}[ht]%
\begin{center}
\includegraphics[width=6cm]{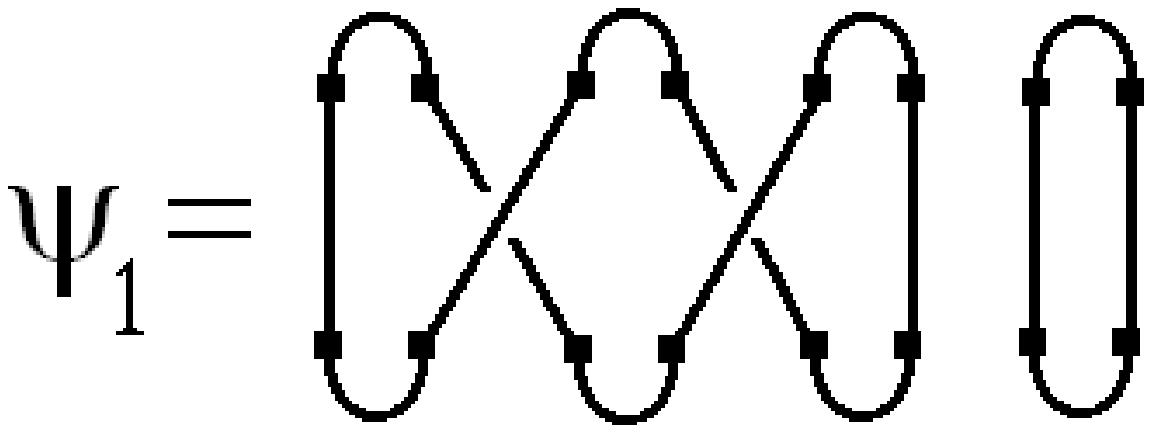}%
\caption{$(\psi_1)^{i_1i_1i_2i_2i_3i_3i_4i_4}_{j_1j_1j_2j_2j_3j_3j_4j_4}=N^2$ for $\psi_1 = (2,3)(4,5) $.}%
\label{cycle3}%
\end{center}
\end{figure}

In figure \ref{cycle2} we show the computation for $\psi_1=(4,5)$ and in figure \ref{cycle3} for $\psi_1=(2,3)(4,5)$.

\begin{figure}[ht]%
\begin{center}
\includegraphics[width=2cm]{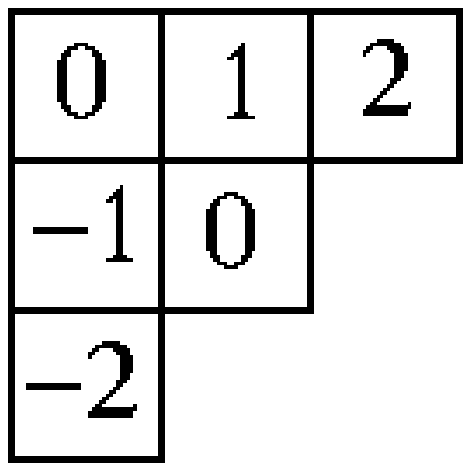}%
\caption{The content of a Young diagram.}%
\label{content}%
\end{center}
\end{figure}

Clearly $(\psi_1)^{i_1 i_1 \cdots i_p i_p}_{j_1 j_1 \cdots j_p j_p}$ gives $N$ to some power. From the examples above we
know that, for example, when $\psi_1=(4,5)=(1,1)(3,3)(4,5)(7,7)$ we have
\bea
 \chi_{(2p)}(\psi_1)
          (\psi_1)^{i_1 i_1 \cdots i_p i_p}_{j_1 j_1 \cdots j_p j_p} = \chi_{(2p)}(\,(4,5)\,)N^3
\eea
and when $\psi_1=(2,3)(4,5)=(1,1)(2,3)(4,5)(7,7)$ we have
\bea
 \chi_{(2p)}(\psi_1)
          (\psi_1)^{i_1 i_1 \cdots i_p i_p}_{j_1 j_1 \cdots j_p j_p} = \chi_{(2p)}(\,(2,3)(4,5)\,)N^2
\eea
Using the convention that $\chi_{(2p)}(\psi+\psi')\equiv \chi_{(2p)}(\psi)+\chi_{(2p)}(\psi')$ we can write
\bea
I &=&2^p p!\sum_{\psi_1\in {\cal B}_p}\chi_{(2p)}(\psi_1)
          (\psi_1)^{i_1 i_1 \cdots i_p i_p}_{j_1 j_1 \cdots j_p j_p}\cr
 &=&2^p p! {\rm Tr}_{(2p)}\left( \prod_{j=0}^{p-1}[N+ \sum_{i=1}^{2j}(i,2j+1) ] \right)
\eea
The sum
\bea
  J_{2j+1}=\sum_{i=1}^{2j}(i,2j+1)
\eea
is known as a Jucys-Murphy element.
These elements form a maximal commutative subalgebra of the symmetric group. 
States in any given irrep can be labeled by standard Young tableau. 
For example, a state in the ${\tiny \yng(3,2,1)}$ irrep is
\bea
  \young(123,45,6)
\eea
We can associate a {\it content} with each box in the Young diagram. 
The box in the upper most row and left most column has content $0$. 
The content for any other specific box is obtained by walking from this upper most and left most box to the 
specific box, adding 1 to the content for every step to the right and subtracting one from the content for 
every step down. 
The content for ${\tiny \yng(3,2,1)}$ is shown in figure \ref{content}.

The Jucys-Murphy element $J_i$ acting on a state labeled by a Young tableau gives the content of box $i$. 
Thus, for example, we have
\bea
  J_2 \young(123,45,6) = \young(123,45,6)\qquad
  J_6 \young(123,45,6) = -2 \young(123,45,6)\qquad
  J_5 \young(123,45,6) = 0  
\eea
It is now a simple matter so see that
\bea
I = \sum_{\sigma\in S_{2p}}\chi_{(2p)}(\sigma)\sigma^{i_1 i_1 \cdots i_p i_p}_{j_1 j_1 \cdots j_p j_p}
  = 2^p p! \prod_{i=1}^{p-1}(N+2i)
\eea

We can also obtain the correlator (\ref{exactvector}) by using Schwinger-Dyson equations. 
We will describe this computation as it provides a nice check of the above representation theoretic computation. 
Recall that for the free vector model we study we have
\bea
 \langle {\cal O}\rangle = \int [dz^k dz_k] \, {\cal O} \, e^{-\, z^i z_i}
\eea
where $z^i=x^i+iy^i$, $z_i=x^i-iy^i$, $z^i z_i\equiv \sum_{i=1}^N z^i z_i$ and
\bea
 [dz^k dz_k] =\prod_{i=1}^N {dx^i \, dy^i\over \sqrt{\pi}}
\eea
Note that
\bea
{\partial z^i\over \partial z^j}=\delta^i_j\qquad
{\partial z_i\over \partial z_j}=\delta_i^j\qquad
{\partial z^i\over \partial z_j}=0
\eea
The first Schwinger-Dyson equation we use is obtained from (repeated indices are summed, regardless of whether they are
up or down)
\bea
 0= \int  [dz^k dz_k] {\partial\over\partial z^i} 
      \left( z^i (z^j z^j)^p(z_m z_m)^p e^{-\, z_l z^l }\right)
\eea
which implies
\bea
(N+2p)\langle (z^i z^i)^p(z_j z_j)^p\rangle =
\langle z^i z_i\, (z^j z^j)^p(z_m z_m)^p\rangle
\eea
The second Schwinger-Dyson equation we use is obtained from
\bea
 0= \int  [dz^k dz_k] {\partial\over\partial z_i} 
      \left( z^i (z^j z^j )^p(z_m z_m)^{p+1} e^{-\, z_l z^l }\right)
\eea
which implies
\bea
 2(p+1)\langle z^i z_i\, (z^j z^j)^p(z_m z_m)^p\rangle =
\langle (z^i z^i)^{p+1} (z_j z_j)^{p+1} \rangle
\eea
Combining these two Schwinger-Dyson equations we find the following recursive relation
\bea
  \langle (z^i z^i)^{p+1} (z_j z_j)^{p+1} \rangle = 2(p+1)(N+2p)\langle (z^i z^i)^p(z_j z_j)^p\rangle
\eea
Using $\langle 1\rangle =1$ we now easily recover (\ref{exactvector}).

\section{Product of Projectors}\label{projprod}

In the derivation of our two point functions we have freely used the product of projection operators in various
places. When the projection operators do not have the same range (this product is needed in equation (\ref{nonstdprojprod}))
the product does not give rise to another projector of the same type. For this reason, in this case, we have spelled out in detail how
the product is computed and what the result is. When the projectors have the same range, the computation is quite standard.
We review this case below.  

For a product of projectors from $V^{\otimes\, q}$ to some irrep of $S_q$ we have
\bea
\cP_R\cP_S &=&\left({1\over q!}\right)^2\sum_{\sigma_1\in S_q}\sum_{\sigma_2\in S_q}\chi_R(\sigma_1)\chi_S(\sigma_2)\sigma_1\sigma_2\cr
&=&\left({1\over q!}\right)^2\sum_{\sigma_1\in S_q}\sum_{\psi\in S_q}\chi_R(\sigma_1)\chi_S(\sigma_1^{-1}\psi)\psi\cr
&=&\left({1\over q!}\right)^2\sum_{\psi\in S_q}{q!\over d_R}\delta_{RS}\chi_R(\psi)\psi\cr
&=&{\delta_{RS}\over d_R}\cP_R
\eea
There is a trivial generalization to the case of projecting from any domain to the irrep of any group ${\cal G}$, derived
exactly as we did above. The projectors are
\bea
P_R ={1\over |{\cal G}|}\sum_{\s\in{\cal G}}\chi_R(\s)\s
\eea
where now $\sigma$ stands for the action on the domain we consider, not necessarily $V^{\otimes q}$ and $\chi_R(\s)$ is the character
of group element $\s$ in irrep $R$ of group ${\cal G}$. 
Use $d_R$ to denote the dimension of this irrep $R$.
The more general product is
\bea
 P_R P_S = {\delta_{RS}\over d_R} P_R
\eea

\section{An important matrix element}\label{ivanovstuff}

In this Appendix we will review the results of \cite{ivanov}. 
We work within the carrier space of some irrep $R$ of $S_{2n}$. 
As above we take $n$ to be even.
Restricting to a $S_n[S_2]$ subgroup, a number of irreps of the subgroup are subduced. 
We are particularly interested in two representations. 
The first representation (that we have
denoted $[S]$ above) has a one dimensional carrier space spanned by the vector $|\eta\rangle$ which obeys
\bea
  \Gamma_R(\sigma)|\eta\rangle =|\eta\rangle \qquad \forall \,\, \sigma\in S_n[S_2]
\eea
The second representation (that we have
denoted $[A]$ above) has a one dimensional carrier space spanned by the vector $|\zeta\rangle$ which obeys
\bea
  \Gamma_R(\sigma)|\zeta\rangle ={\rm sgn}(\sigma)|\zeta\rangle \qquad \forall \,\, \sigma\in S_n[S_2]
\eea
Only representations $R$ corresponding to Young diagrams with both an even number of rows and an even number of columns
subduce both $[S]$ and $[A]$.
In this case, the number of boxes in Young diagram $R\vdash 2n$ is a multiple of 4, so that $n$ is indeed even.
Let $\nu\vdash {n\over 2}$. 
Denote the parts of $\nu$ by $(\nu_1,\nu_2,...)$ and the number of parts of $\nu$ by $l(\nu)$. 
For example, $\nu=(4,4)$ is a partition of 8 with two parts, $\nu_1=\nu_2=4$ and $l(\nu)=2$.
Using $\nu$ we can construct the cycle
\bea
  \sigma_{4\nu}=(1,2,...,4\nu_1)(4\nu_1+1,4\nu_1+2,...,4\nu_1+4\nu_2)\cdots (2n-2\nu_k+1,\cdots,2n)
\eea
Ivanov\cite{ivanov} has computed the matrix element
\bea
   \langle \eta |\Gamma_R(\sigma_{4\nu})|\zeta\rangle ={2^{l(\nu)}\sqrt{{\rm hooks}_R}\over n!2^n}\chi_{R/4}(\nu)
\eea

For the operators in (\ref{opone}), to compute the two point function we need to consider $\nu$ the identity so that $l(\nu)={n\over 2}$. 
Thus, the matrix element we need is
\bea
   \langle \eta |\Gamma_R(\sigma_{4\nu})|\zeta\rangle ={2^{n\over 2}\sqrt{{\rm hooks}_R}\over n!2^n}d_{R/4}
\eea

For the operators in (\ref{oldopone}), to compute the two point function we need to consider $\nu$ an ${n\over 2}$ cycle so that $l(\nu)=1$. 
Thus, the matrix element we need is
\bea
   \langle \eta |\Gamma_R(\sigma_{4\nu})|\zeta\rangle ={2\sqrt{{\rm hooks}_R}\over n!2^n}\chi_{R/4}(\nu)
\eea
Using well know properties of the character of an ${n\over 2}$ cycle in group $S_{n\over 2}$, we know that 
the right hand side is only non-zero when $R/4$ is a hook and in this case 
\bea
   \langle \eta |\Gamma_R(\sigma_{4\nu})|\zeta\rangle =\pm {2\sqrt{{\rm hooks}_R}\over n!2^n}
\eea
We do not spell out the sign on the right hand side which depends on $R/4$ because it is not needed - we 
only use the square of this matrix element.

\section{Some Examples}

To test our formula (\ref{mainresult}) we have computed some examples of our polynomials explicitely. 
It is then rather straight forward to verify the correctness of (\ref{mainresult}) by simply using the 
free field Wick contractions. 
In all cases we have studied, (\ref{mainresult}) is correct.
 
Some examples of our operators are
\bea
  O_{\tiny \yng(2,2)}(Z)    ={1\over 2}{\rm Tr}(Z^2)
\eea
\bea
  O_{\tiny \yng(4,4)}(Z)    = {1\over 28}{\rm Tr}(Z^2)^2+{1\over 14}{\rm Tr}(Z^4)
\eea
\bea
  O_{\tiny \yng(2,2,2,2)}(Z)= {1\over 28}{\rm Tr}(Z^2)^2-{1\over 14}{\rm Tr}(Z^4)
\eea
Operators involving 6 fields involve sums over $S_{12}$, which are beyond 
the reach of our numerical methods.
The corresponding two point functions are
\bea
\langle O_{\tiny\yng(2,2)}O^\dagger_{\tiny\yng(2,2)}\rangle = N(N-1)
\eea
\bea
\langle O_{\tiny\yng(4,4)}O^\dagger_{\tiny\yng(4,4)}\rangle ={4\over 49}N(N-1)(N+1)(N+2)
\eea
\bea
\langle O_{\tiny\yng(4,4)}O^\dagger_{\tiny\yng(2,2,2,2)}\rangle = 0
\eea
\bea
\langle O_{\tiny\yng(2,2,2,2)}O^\dagger_{\tiny\yng(2,2,2,2)}\rangle = {4\over 49}N(N-1)(N-2)(N-3)
\eea
We have computed the right hand side of these expressions using our formula (\ref{mainresult}), 
using the operators given above and doing the free field Wick rotations and finally by evaluating the
sum (\ref{twpnt}) directly. All three answers are in complete agreement in all cases.

\end{appendix}

\end{document}